\documentclass{article}

\PassOptionsToPackage{numbers, compress}{natbib}
\usepackage[preprint]{neurips_2026}

\usepackage[utf8]{inputenc} 
\usepackage[T1]{fontenc}    
\usepackage[hidelinks]{hyperref}       
\usepackage{url}            
\usepackage{booktabs}       
\usepackage{amsfonts}       
\usepackage{bm}
\usepackage{nicefrac}       
\usepackage{microtype}      
\usepackage{xcolor}         
\usepackage{graphicx}       
\usepackage{listings}
\usepackage{color}
\usepackage[acronym,nomain,nonumberlist,nohypertypes={acronym}]{glossaries}
\usepackage{cleveref}       
\makeglossaries
\newacronym{sharp}{SHARP}{Scientific Human-Agent Reproduction Pipeline}

\definecolor{dkgreen}{rgb}{0,0.6,0}
\definecolor{gray}{rgb}{0.5,0.5,0.5}
\definecolor{mauve}{rgb}{0.58,0,0.82}

\title{A Scientific Human-Agent Reproduction Pipeline}

\author{%
  Joschka Birk\\
  Universit\"at Hamburg\\
  \texttt{joschka.birk@uni-hamburg.de} \\
  \And
  Gregor Kasieczka\\
  Universit\"at Hamburg\\
  \texttt{gregor.kasieczka@uni-hamburg.de} \\
  \And
  Siddharth Mishra-Sharma\thanks{Also at Anthropic}\\
  Boston University\\
  \texttt{smishras@bu.edu}\\
  \And
  Benjamin Nachman\\
  Stanford University\\
  SLAC National Accelerator Laboratory\\
  \texttt{nachman@stanford.edu} \\
  \And
  Dennis Noll\\
  Stanford University\\
  SLAC National Accelerator Laboratory\\
  \texttt{nollde@stanford.edu} \\
  \And
  Tanvi Wamorkar\\
  Stanford University\\
  SLAC National Accelerator Laboratory\\
  \texttt{twamorka@stanford.edu}
}

\begin{document}

\maketitle

\begin{abstract}
Reproducing scientific analyses is essential for preserving knowledge, building extensible codebases, and deepening researcher understanding -- yet the effort often outweighs its academic recognition.
We argue that the reproduction of scientific data analyses is fundamentally a translation task: converting human-readable knowledge (papers, documentation) into machine-readable analysis code.
This makes it uniquely well-suited for AI agents.
We present SHARP (Scientific Human-Agent Reproduction Pipeline), a structured framework for reproducing scientific analyses through human-agent collaboration.
SHARP decomposes a reproduction task into discrete steps, which an AI agent executes autonomously using specialized subagents for code generation, testing, and quality assurance.
At defined checkpoints, the researcher reviews progress, provides feedback, and steers the analysis - keeping the human firmly in control of scientific judgment while the agent handles implementation.
We demonstrate SHARP by reproducing a jet classification task in particle physics from a published paper.
We evaluate the reproduction along three axes: analysis performance against the original results, code quality and faithfulness, and the nature of the human-agent conversation.
The latter is evaluated with a novel framework for characterizing human-agent interactions.
Our work highlights a practical model for AI-assisted scientific reproduction where the researcher's role shifts from writing code to understanding, evaluating, and directing -- elevating human understanding rather than replacing it.
\end{abstract}

\section{Introduction}

Reproducing a published scientific data analysis yields two things of lasting value: an extensible codebase that preserves the analysis and can be adapted to explore related problems, and a deep researcher understanding of the processes, assumptions, and pitfalls that lie between the analysis methods and the numerical results.
Despite these benefits, the effort required to reproduce an analysis often outweighs its academic recognition, and, as a consequence, published results are rarely independently reproduced.
Modern AI coding agents offer a path to make this effort substantially easier.

The reproduction of data analyses is particularly well suited to AI agents because it is a \emph{translation} task, not an \emph{extrapolation} task.
A paper describing such an analysis is ideally a self-sufficient specification of the process.
Reproduction then amounts to converting this human-readable specification into machine-readable code.
This framing casts the agent as a meticulous translator rather than a creative problem-solver.
With the agent handling the translation, reproduction becomes faster and easier, and the researcher can devote their time to scientific judgment -- reading, testing, and understanding the resulting analysis.
An agent-driven workflow can also enforce consistent code standards and quality across analysis reproductions, without requiring different researchers to internalize best practices from scratch.

Recent work explores a variety of different approaches to agents in scientific research~\cite{schwartz_agentic,shih_agentic_a,shih_agentic_b,ai_scientist,haichen_agentic,autoresearch,anon_agentic,prbench,Plehn:2026gxv,Qiu:2026iby,liu2026scifi}.
We present \gls{sharp}, a specialized agent harness and collaborative workflow for the guided reproduction of scientific analyses, jointly run by a human and an agent.
During the reproduction process, the agent, which is built on Claude Code, has access to the paper, its supporting documents, and selected online resources.
The agent and the human first agree on a reproduction plan; the agent then works on subtasks autonomously, interacting with the human at predefined checkpoints.
With this structure, \gls{sharp} occupies a deliberate middle ground between fixed, task-specific pipelines such as Agents of Discovery~\cite{agents_of_discovery} or \texttt{CoLLM}~\cite{collm} and more free-form agent workflows such as the Just Furnish Context~\cite{jfc}: structured enough to enforce consistency and quality, yet flexible enough to tackle non-trivial, real-world scientific analyses.

\section{The SHARP Framework}
\begin{figure}
  \centering
  \includegraphics[width=0.99\linewidth]{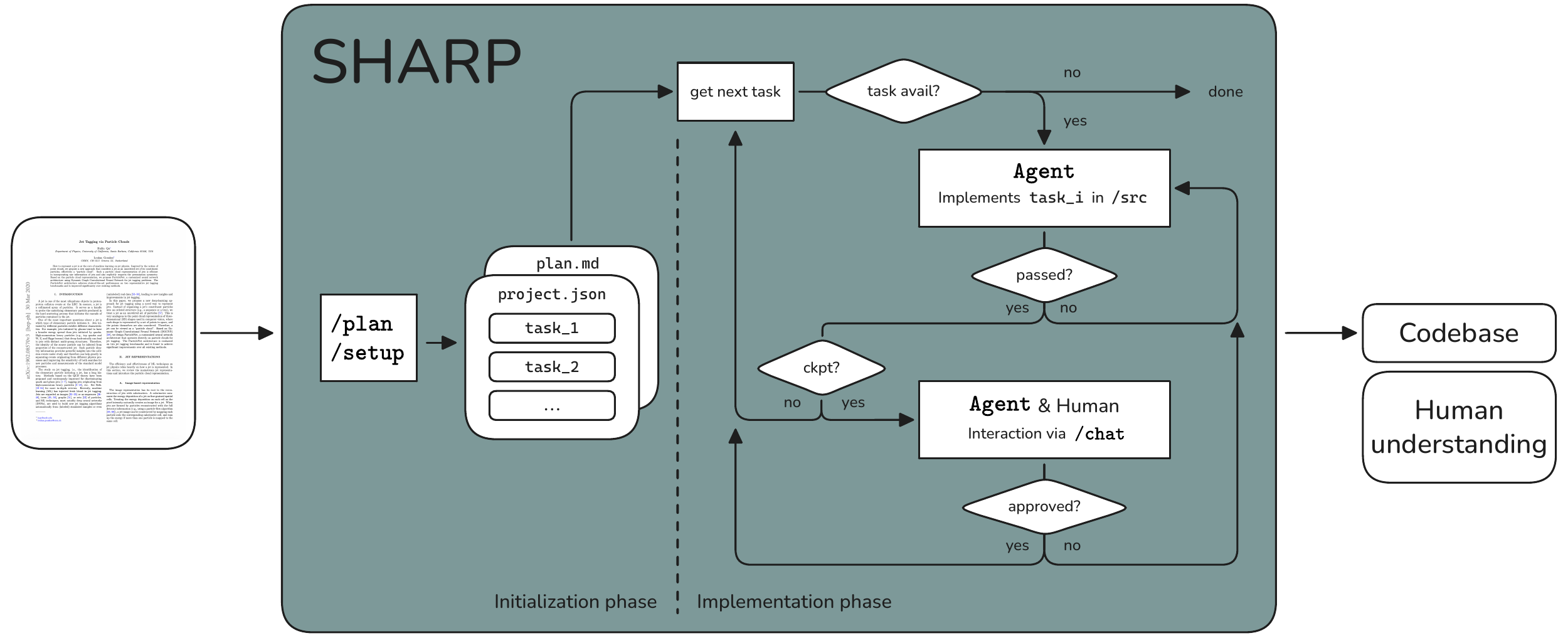}
  \caption{Starting from an initial input -- including the paper to be reproduced -- \gls{sharp} first produces a reproduction plan, then iteratively solves the individual tasks of the plan, with the agent working autonomously and the human intervening at checkpoints. The final output is a reproduced codebase and a deepened human understanding of the analysis.}
  \label{fig:workflow}
\end{figure}

The \gls{sharp} workflow, illustrated in \Cref{fig:workflow}, is iterative and largely follows Geoffrey Huntley's Ralph pattern~\cite{ralph}.
A user starts from a \gls{sharp} template~\cite{sharp} that provides the basic project structure, software environment, and agent setup.
The workflow is built on Claude Code \texttt{v2.1.92}~\cite{claude_code} and uses \texttt{claude-opus-4.6} as the underlying LLM.

Invoked via the \texttt{/plan} skill, the agent first studies the paper in question and asks clarifying questions to understand the task and the user's expectations.
Then, it proposes a decomposition of the reproduction into a set of individual tasks (ten by default), some of which are marked as checkpoints, stored in \texttt{plan.md}.
The user can refine or redirect the plan, including the checkpoints, through further discussion with the agent.
Once the plan is approved, the agent compiles it into a machine-readable project specification, \texttt{project.json}, via the \texttt{/setup} skill.
This project specification drives the subsequent execution phase.

In each iteration, the agent solves a single task from the plan, which reduces the needed context window and allows the process to be paused and resumed at any task boundary.
State is handed off between iterations via \texttt{git} version control, a small set of project files (\texttt{project.json}, \texttt{progress.txt}) that record the current plan and the progress so far, and the state of the codebase.
Within each iteration, the agent delegates to specialized subagents in parallel: a \emph{Paper Analyst} extracts information from the paper, \emph{Code} and \emph{Test} subagents implement the task in a test-driven style, a \emph{Statistician} handles statistical analysis, and a \emph{Critic} enforces the \emph{FlexCAST} principles of modularity, testability, and robustness~\cite{flexcast}.
Modularity is implemented through the \texttt{law}~\cite{Rieger:2024tkw} workflow engine, and the generated code runs in a self-contained Conda environment set up, along with other project configuration, by a single \texttt{setup.sh} script.
The agent explicitly evaluates whether the task is complete based on the tests and the project specification.

If the completed task is marked as a checkpoint, the agent pauses execution and hands control back to the user for review.
To streamline this step, the agent summarizes the work it has just completed and advises the user on how to run the code, execute the relevant tests, and spot likely issues.
Through the \texttt{/chat} interface, the user can discuss the solution in more detail before either approving the implementation or requesting revisions.
Once the checkpoint is cleared, the agent proceeds to the next task and the process continues until all tasks are completed.

\section{Experimental Setup}
At the Large Hadron Collider at CERN, high-energy proton collisions produce unstable particles whose decay products can form \emph{jets} -- collimated sprays of detectable particles.
\emph{Jet tagging} is the task of classifying jets by the type of particle that initiated them, and is a central ML task in particle physics.
We focus on the reproduction of the ParticleNet~\cite{Qu:2019gqs} paper, which introduces a graph neural network that treats jets as point clouds, evaluated on top quark vs.\ QCD jet classification using a public dataset~\cite{kasieczka_2019_2603256,Kasieczka:2019dbj}.
The reproduction targets four metrics for the ParticleNet-Lite model reported in~\cite{Qu:2019gqs}: accuracy, area under the ROC curve (AUC), and background rejection at 50\% and 30\% signal efficiency ($R_{50}$ and $R_{30}$).
We choose the ParticleNet-Lite architecture as it is a non-trivial architecture for jet tagging while requiring only moderate resources.
To obtain the target metrics, the agent is expected to implement multiple non-trivial stages including dataset download and preparation, model training and evaluation, and uncertainty estimation.
The reproduction starts with a prompt from the human (Prompt~\ref{plan_prompt}), specifying the paper, target metrics, and engineering requirements.

\begin{lstlisting}[caption={Prompt used to initiate the replication of the ParticleNet-Lite results.}, label={plan_prompt}]
Load the plan skill and create a PRD for replication (incl reimplementation) of arXiv:1902.08570.
- Running in a container with 1 GPU. Ensure that the code runs on GPU, but also add CPU support as fallback.
- Only ParticleNet-Lite, only top tagging. Reproduce accuracy, AUC, and 1/eps_b results.
- Provide smoke-test mode (`--n-events N`) that runs train->evaluate->plot on a tiny subset in a short time.
- Use PyTorch + PyTorch Lightning. Create validation plots of input feature distributions vs physical expectations.
- Versioned output (`--version i`) creating separate results folders. Plots in PDF.
- Ensure that uncertainties on the final results are handled exactly as in the paper.
\end{lstlisting}

Both the agent and the human operate inside \texttt{claude-hpc}, a sandboxed Claude Code environment specifically developed for this project~\cite{claude-hpc}.
\texttt{claude-hpc} executes Claude Code in an isolated container on HPC systems -- in our case NERSC's Perlmutter.
Inside the container, file system access is restricted to a set of explicitly allowed directories, and network access is limited to a curated allowlist of approved services including GitHub, arXiv, and Conda.
In particular, the agent has access to public code repositories associated with the target paper, and has been observed to occasionally consult them.
Furthermore, the container has access to a single NVIDIA A100 GPU on which all training runs are performed.

\section{Results}

Table~\ref{tab:results} shows the results from three independent \gls{sharp} runs of the reproduction task compared to the values reported in the original paper. All three runs were performed under a flat-rate \$200/month Claude Max subscription, with no per-token API charges.

For each \gls{sharp} run we report the median and spread of 9 classifier trainings. 
In all runs, the performance agrees with the published ParticleNet-Lite result, except for slight deviations in the $R_{30}$ metric, which is particularly sensitive. This shows that most of the results are well reproduced with hints to minor differences to the official publication.

\begin{table}[h]
    \centering
    \caption{
        Reproduced ParticleNet-Lite~\cite{Qu:2019gqs} performance compared to the original paper.
        We report the performance of the training with the median-accuracy.
        Uncertainties on the accuracy and AUC are not compared as they are not reported in the original paper.
        Uncertainties on the background rejection at signal efficiency equal to 50\% ($R_{50}$) and 30\% ($R_{30}$) are obtained from the standard deviation over all classifier trainings.
    }
    \begin{tabular}{lcccc}
        \toprule
          & \textbf{Accuracy} & \textbf{AUC} & $\bm{R_{50}}$ & $\bm{R_{30}}$ \\
        \midrule
        ParticleNet-Lite (paper)  & $0.937$ & $0.9844$ & $325 \pm 5$ & $1262 \pm 49$ \\
        \gls{sharp} Run 1      & $0.938$ & $0.9844$ & $318 \pm 7$ & $1278 \pm 41$ \\
        \gls{sharp} Run 2      & $0.938$ & $0.9844$ & $317 \pm 7$ & $1188 \pm 41$ \\
        \gls{sharp} Run 3      & $0.938$ & $0.9845$ & $323 \pm 6$ & $1147 \pm 50$ \\
        \bottomrule
    \end{tabular}
    \label{tab:results}
\end{table}

The agent produces a well-structured \texttt{law}-based workflow, giving the human a legible entry point into the generated codebase. The generated codebase contains $\mathcal{O}(100)$ unit tests with 99\% \texttt{pytest} line coverage of production code (not including \texttt{law} or plotting code).
In addition, individual model checkpoints from the \gls{sharp} runs were independently validated by an external evaluation script written by a human expert, using only the plain PyTorch model definition and the corresponding weights.

To characterize the human-agent conversation, we use \texttt{claude-parser}~\cite{claude-parser}, a tool specifically developed for this work.
For each human message, the tool calls \texttt{claude-haiku-4-5} with the full conversation as context, and asks it to assign a \emph{complexity} and a \emph{type}.
\emph{Complexity} captures how much valuable information the message contributes to the reproduction -- that is, how much worse off the agent would be without it: \emph{easy} (approvals, error reports), \emph{medium} (a concrete parameter, dataset, or correction), or \emph{hard} (dense guidance, methodological insight, or scope-defining input).
\emph{Type} captures the role of the intervention: \emph{essential} (required for faithfulness to the paper), \emph{optional} (extra scope beyond the paper), or \emph{meta} (workflow steering, e.g.\ ``commit now'').

Table~\ref{tab:interactions} reports the resulting distribution of human messages across the three reproduction runs.
In every run, the first human message -- specifying the reproduction target and engineering requirements -- is classified as a \emph{hard essential} intervention, providing the agent with the scope and evaluation targets of the reproduction.
Run 1 proceeded smoothly with only nine total human messages, while runs 2 and 3 required additional steering that is dominated by medium-complexity essential interventions (concrete corrections and clarifications) and easy meta messages for workflow actions such as commits.
Each reproduction run took roughly one working day from the initial prompt to the final results, with the user operating two runs in parallel at medium intensity -- attending to the agent between other unrelated tasks throughout the day.

\begin{table}[h]
    \centering
    \caption{Distribution of human messages across the three reproduction runs, classified by \emph{type} (\emph{essential}, \emph{optional}, \emph{meta}) and \emph{complexity} (H: hard, M: medium, E: easy). The first human message -- specifying the reproduction target and engineering requirements (Prompt 1) -- is hard essential by construction in every run, as it defines the scope and evaluation targets of the reproduction. Remaining messages reflect organic human-agent interaction during execution.}
    \label{tab:interactions}
    \begin{tabular}{lccc}
        \toprule
        \textbf{Run} & \textbf{Essential (H/M/E)} & \textbf{Optional (H/M/E)} & \textbf{Meta (H/M/E)} \\
        \midrule
        \gls{sharp} Run 1 ($n=9$)  & 1 / 1 / 1   & 1 / 0 / 0 & 0 / 0 / 5  \\
        \gls{sharp} Run 2 ($n=24$) & 3 / 6 / 1   & 0 / 1 / 1 & 0 / 0 / 12 \\
        \gls{sharp} Run 3 ($n=31$) & 2 / 13 / 0  & 0 / 2 / 1 & 0 / 1 / 12 \\
        \bottomrule
    \end{tabular}
\end{table}

\paragraph{Limitations}

The agent occasionally introduces subtle implementation differences from the paper -- such as the learning rate schedule, activation functions, or aggregation choice -- which typically have only minor impact on the final metrics and are usually caught by repeated cross-checking prompts.
A more severe failure mode concerns dataset-specific pitfalls that require domain knowledge: the public top tagging dataset, for instance, stores a truth-label particle that trivially invalidates the classification task if inadvertently loaded.
This failure did not occur in our reported runs but was encountered during development and went undetected by automated tests. The three reported runs comprise all runs performed with the published setup; no run was abandoned or failed to complete with human-in-the-loop steering. The degree of human input required in each run is reflected in the conversation analysis (Table~\ref{tab:interactions}).
Finally, our evaluation rests on a single reproduction; how well \gls{sharp} generalizes across analyses and domains is left for future work.

\section{Discussion and Conclusion}

This work has introduced \gls{sharp}, a structured framework for reproducing scientific analyses through human-agent collaboration, and demonstrated its effectiveness via the reproduction of a non-trivial result from particle physics: a graph-based neural network for jet classification on point clouds.
We report on the resulting performance, code quality, and human-agent interaction.

The reproduced results match the original paper with high precision (within 0.1 percentage points in accuracy). Human-agent interaction varied across runs, from highly efficient exchanges to runs requiring more substantive methodological clarification.
SHARP handles codebase preparation and workflow setup with high fidelity. Correctly reproducing finer implementation details requires targeted prompting and iterative cross-checking, and remains an important limitation for scientific contexts. A key latent risk is domain-specific failure modes that automated tests cannot catch, such as truth-label leakage, underscoring that human oversight remains essential.

We see analysis reproduction as a high-value target for agentic AI.
Benchmarks are abundant, and reproduction both deepens researcher understanding and provides a foundation for downstream innovation.
Promising directions for future work include improved agent self-consistency and a more structured test design.

\section*{Code Availability}

The \gls{sharp} template is available at: \url{https://github.com/stanford-ai4physics/sharp}.
The \texttt{claude-parser} tool for the conversation analysis is available at: \url{https://github.com/nollde/claude-parser}.
The \texttt{claude-hpc} containment tool is available at: \url{https://github.com/nollde/claude-hpc}.

\section*{Acknowledgments}
This project originated at the ``Human Meets AI in Scientific Research Replication'' hackathon at the Center for Decoding the Universe at Stanford University -- we would like to thank the organizers, especially Ioana Ciuc\u{a}.
B.N., D.N., and T.W. are supported by the Department of Energy (DOE), Office of Science under contract DE-AC02-76SF00515.
T.W. is supported by the National Science Foundation under Grant No. 2311666.
This research used resources of the National Energy Research Scientific Computing Center, a DOE Office of Science User Facility supported by the Office of Science of the U.S. Department of Energy under Contract No. DE-AC02-05CH11231 using NERSC award HEP-ERCAP0035546.
J.B. and G.K. are supported by the DFG under the German Excellence Initiative -- EXC 2121  Quantum Universe – 390833306.
J.B. is supported by a scholarship of the German Academic Exchange Service~(DAAD).
J.B. also acknowledges support via the Hamburg VISTA/VISOR -- Virtual Initiative for Science \& Technology in AI -- network.

\newpage

\bibliography{refs}

@article{Qu:2019gqs,
    author = "Qu, Huilin and Gouskos, Loukas",
    title = "{{ParticleNet: Jet Tagging via Particle Clouds}}",
    eprint = "1902.08570",
    archivePrefix = "arXiv",
    primaryClass = "hep-ph",
    doi = "10.1103/PhysRevD.101.056019",
    journal = "Phys. Rev. D",
    volume = "101",
    number = "5",
    pages = "056019",
    year = "2020"
}

@misc{agents_of_discovery,
      title={{Agents} of {Discovery}}, 
      author={Sascha Diefenbacher and Anna Hallin and Gregor Kasieczka and Michael Krämer and Anne Lauscher and Tim Lukas},
      year={2026},
      eprint={2509.08535},
      archivePrefix={arXiv},
      primaryClass={hep-ph},
      url={https://arxiv.org/abs/2509.08535}, 
}

@misc{liu2026scifi,
  title={SciFi: A Safe, Lightweight, User-Friendly, and Fully Autonomous Agentic AI Workflow for Scientific Applications},
  year={2026},
  author={Liu, Qibin and Gonski, Julia},
  eprint={2604.13180},
  archivePrefix={arXiv},
  primaryClass={cs.AI},
  url={https://arxiv.org/abs/2604.13180}
}

@misc{jfc,
      title={{AI Agents Can Already Autonomously Perform Experimental High Energy Physics}}, 
      author={Eric A. Moreno and Samuel Bright-Thonney and Andrzej Novak and Dolores Garcia and Philip Harris},
      year={2026},
      eprint={2603.20179},
      archivePrefix={arXiv},
      primaryClass={hep-ex},
      url={https://arxiv.org/abs/2603.20179}, 
}

@misc{schwartz_agentic,
      title={{Resummation of the C-Parameter Sudakov Shoulder Using Effective Field Theory}}, 
      author={Matthew D. Schwartz},
      year={2026},
      eprint={2601.02484},
      archivePrefix={arXiv},
      primaryClass={hep-ph},
      url={https://arxiv.org/abs/2601.02484}, 
}

@misc{shih_agentic_a,
      title={{Learning to Unscramble: Simplifying Symbolic Expressions via Self-Supervised Oracle Trajectories}}, 
      author={David Shih},
      year={2026},
      eprint={2603.11164},
      archivePrefix={arXiv},
      primaryClass={hep-th},
      url={https://arxiv.org/abs/2603.11164}, 
}

@misc{shih_agentic_b,
      title={{Learning to Unscramble Feynman Loop Integrals with SAILIR}}, 
      author={David Shih},
      year={2026},
      eprint={2604.05034},
      archivePrefix={arXiv},
      primaryClass={hep-ph},
      url={https://arxiv.org/abs/2604.05034}, 
}

@misc{ai_scientist,
      title={{The AI Scientist: Towards Fully Automated Open-Ended Scientific Discovery}}, 
      author={Chris Lu and Cong Lu and Robert Tjarko Lange and Jakob Foerster and Jeff Clune and David Ha},
      year={2024},
      eprint={2408.06292},
      archivePrefix={arXiv},
      primaryClass={cs.AI},
      url={https://arxiv.org/abs/2408.06292}, 
}

@misc{haichen_agentic,
      title={{Automating} {High} {Energy} {Physics} {Data} {Analysis} with {LLM-Powered} {Agents}}, 
      author={Eli Gendreau-Distler and Joshua Ho and Dongwon Kim and Luc Tomas Le Pottier and Haichen Wang and Chengxi Yang},
      year={2025},
      eprint={2512.07785},
      archivePrefix={arXiv},
      primaryClass={physics.data-an},
      url={https://arxiv.org/abs/2512.07785}, 
}

@misc{anon_agentic,
      title={{Agentic AI -- Physicist Collaboration in Experimental Particle Physics: A Proof-of-Concept Measurement with LEP Open Data}}, 
      author={Anthony Badea and Yi Chen and Marcello Maggi and Yen-Jie Lee and Electron-Positron Alliance},
      year={2026},
      eprint={2603.05735},
      archivePrefix={arXiv},
      primaryClass={hep-ex},
      url={https://arxiv.org/abs/2603.05735}, 
}

@misc{autoresearch,
	title = {karpathy/autoresearch},
	url = {https://github.com/karpathy/autoresearch},
	urldate = {2026-04-13},
	author = {Andrej Karpathy},
	month = apr,
	year = {2026},
	note = {original-date: 2026-03-06T22:00:43Z},
}

@misc{prbench,
      title={{PRBench: End-to-end Paper Reproduction in Physics Research}}, 
      author={Shi Qiu and others},
      year={2026},
      eprint={2603.27646},
      archivePrefix={arXiv},
      primaryClass={cs.CL},
      url={https://arxiv.org/abs/2603.27646}, 
}

@misc{ralph,
	author = "Huntley, Geoffrey",
	title = {{Ralph} {Wiggum} as a "software engineer"},
	url = {https://ghuntley.com/ralph/},
	language = {en},
	urldate = {2026-04-13},
	journal = {Geoffrey Huntley},
	month = jul,
	year = {2025},
}

@article{Rieger:2024tkw,
    author = "Rieger, Marcel",
    title = "{{End-to-End Analysis Automation over Distributed Resources with Luigi Analysis Workflows}}",
    eprint = "2402.17949",
    archivePrefix = "arXiv",
    primaryClass = "physics.data-an",
    doi = "10.1051/epjconf/202429505012",
    journal = "EPJ Web Conf.",
    volume = "295",
    pages = "05012",
    year = "2024"
}

@misc{flexcast,
    title = "{FlexCAST: Enabling Flexible Scientific Data Analyses}",
    author = "Nachman, Benjamin and Noll, Dennis",
    eprint = "2507.11528",
    archivePrefix = "arXiv",
    primaryClass = "hep-ex",
    year = "2025",
    url={https://arxiv.org/abs/2507.11528},
}

@misc{claude_code,
	title = {Claude {Code} {Docs}},
	url = {https://code.claude.com/docs/en/overview},
	language = {en},
	urldate = {2026-04-14},
	journal = {Claude Code Docs},
	author = {{Anthropic}},
    year = "2026",
}

@misc{sharp,
	title = {{SHARP}: {Template} to reproduce scientific analyses with a coding agent.},
    author = {Noll, Dennis and Birk, Joschka},
    url = {https://github.com/stanford-ai4physics/sharp},
	copyright = {MIT},
	urldate = {2026-04-15},
    year = "2026",
    month = "April",
}

@misc{claude-hpc,
	title = {claude-hpc: {Sandboxed} {Claude} {Code} environment for {HPC} and local {Docker}},
    author = {Noll, Dennis and Birk, Joschka},
    url = {https://github.com/nollde/claude-hpc},
	copyright = {MIT},
	urldate = {2026-04-15},
    year = "2026",
    month = "April",
}

@misc{claude-parser,
	title = {claude-parser: {Display} and {Analyze} {Conversations} with {Claude}},
    author = {Noll, Dennis},
    url = {https://github.com/nollde/claude-parser},
    shorttitle = {nollde/claude-parser},
	urldate = {2026-04-15},
    year = "2026",
    month = "April",
}

@article{Kasieczka:2019dbj,
  author        = {Butter, Anja and others},
  editor        = {Kasieczka, Gregor and Plehn, Tilman},
  title         = {{The Machine Learning landscape of top taggers}},
  eprint        = {1902.09914},
  archiveprefix = {arXiv},
  primaryclass  = {hep-ph},
  doi           = {10.21468/SciPostPhys.7.1.014},
  journal       = {SciPost Phys.},
  volume        = {7},
  pages         = {014},
  year          = {2019}
}

@dataset{kasieczka_2019_2603256,
  author    = {Kasieczka, Gregor and
               Plehn, Tilman and
               Thompson, Jennifer and
               Russel, Michael},
  title     = {{Top Quark Tagging Reference Dataset}},
  month     = mar,
  year      = 2019,
  publisher = {Zenodo},
  version   = {v0 (2018\_03\_27)},
  doi       = {10.5281/zenodo.2603256},
  url       = {https://doi.org/10.5281/zenodo.2603256}
}

@misc{collm,
      title={{CoLLM: AI engineering toolbox for end-to-end deep learning in collider analyses}}, 
      author={W. Esmail and A. Hammad and M. Nojiri},
      year={2026},
      eprint={2602.06496},
      archivePrefix={arXiv},
      primaryClass={hep-ph},
      url={https://arxiv.org/abs/2602.06496}, 
}

@misc{Plehn:2026gxv,
    author = "Plehn, Tilman and Schiller, Daniel and Schmal, Nikita",
    title = "{MadAgents}",
    eprint = "2601.21015",
    url = {https://arxiv.org/abs/2601.21015},
    archivePrefix = "arXiv",
    primaryClass = "hep-ph",
    year = "2026"
}

@misc{Qiu:2026iby,
    author = "Qiu, Shi and Cai, Zeyu and Wei, Jiashen and Li, Zeyu and Yin, Yixuan and Cao, Qing-Hong and Liu, Chang and Luo, Ming-xing and Yuan, Xing-Bo and Zhu, Hua Xing",
    title = "{An End-to-end Architecture for Collider Physics and Beyond}",
    eprint = "2603.14553",
    url = {https://arxiv.org/abs/2603.14553},
    archivePrefix = "arXiv",
    primaryClass = "hep-ph",
    reportNumber = "CPTNP-2026-012",
    year = "2026"
}

\medskip

{
\small

}

\end{document}